\documentclass[11pt,a4paper]{article}
\usepackage{jheppub}

\usepackage{amsmath}
\usepackage[dvipsnames]{xcolor}
\usepackage{graphicx}
\usepackage{hyperref}
\usepackage{mathtools}
\usepackage{graphicx}
\usepackage{comment}
\usepackage{cancel}
\usepackage{physics}
\usepackage{mathtools}
\usepackage[caption=false,justification=justified]{subfig} 
\usepackage{amsfonts}

\hypersetup{
 bookmarksnumbered=true,
 colorlinks=true,
 linkcolor=[rgb]{0.098,0.098,0.439},
 citecolor=[rgb]{0.098,0.098,0.439},
 urlcolor=[rgb]{0.098,0.098,0.439}
}

\title{Oscillating scalar potential and its implications for cosmic neutrino background searches}

\author[a]{Yechan Kim,}
\author[a]{Hye-Sung Lee}
\affiliation[a]{Department of Physics, Korea Advanced Institute of Science and Technology\\ Daejeon 34141, Korea}
\emailAdd{cj7801@kaist.ac.kr}
\emailAdd{hyesung.lee@kaist.ac.kr}

\date{March 2025}

\abstract{
We propose a novel mechanism in which the mass-squared term of a scalar potential periodically switches sign due to an external oscillatory wave. As a result of this ``potential oscillation,'' the vacuum transitions between symmetry-broken and symmetry-restored phases.
This repeated toggling leads to a time-varying vacuum state that, through the scalar field’s couplings to other sectors, gives rise to interesting phenomenological consequences. As a concrete illustration, we demonstrate how these oscillations can open a new avenue for probing the cosmic neutrino background.}

\begin{document}
\maketitle
\flushbottom

\section{Introduction}
\label{IntroSec}

Symmetry and its breaking are central to many areas of physics, from the subatomic scale to phase transitions in condensed matter to cosmological structures. Depending on the shape of the scalar potential, spontaneous symmetry breaking may occur. This phenomenon underlies mass generation in the Standard Model (SM) of particle physics~\cite{Englert:1964et, Higgs:1964pj, Guralnik:1964eu}.

Here, we propose a new mechanism in which the scalar potential oscillates due to coupling with external waves, such as wave dark matter (wave DM). 
The shape of the potential itself oscillates at the frequency of the external wave. 
This ``potential oscillation'' can periodically shift the vacuum state, causing the symmetry to break during one interval of time and be restored during another.\footnote{A time-dependent potential has been investigated in the context of reheating after inflation~\cite{Kofman:1994rk, Kofman:1997yn, Bassett:1997az} and in axion production~\cite{Co:2017mop, Nakayama:2021avl}. Also, symmetry restoration in the early universe via finite temperature effects has been studied~\cite{Kirzhnits:1972ut, Senaha:2020mop, Nakayama:2021avl}.
In particular, Ref.~\cite{Nakayama:2021avl} discusses the scalar mixing between the SM Higgs and a dark Higgs in the context of symmetry breaking in the early universe, taking into account thermal effects.
However, periodic symmetry restoration has not been discussed previously.}

Such an oscillation between symmetry-broken and restored phases yields a remarkably simple yet distinctive phenomenology. 
To illustrate this with a concrete example, we first introduce the mechanism and then focus on the still-mysterious neutrino mass sector in particle physics. If the neutrino’s Majorana mass arises through this mechanism, the resulting modulation could enable novel searches for the cosmic neutrino background (C$\nu$B), whose direct detection is widely regarded as a major milestone in particle physics.

\section{Potential oscillation}
\label{PoOscSec}

\begin{figure}[tb]
    \centering
    \includegraphics[width=0.65\textwidth]{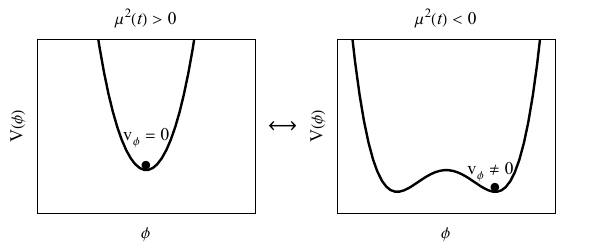}
    \caption{Schematic representation of the potential oscillation. The sign of $\mu^2(t)$ changes periodically, inducing transitions between zero and nonzero VEV of $\phi$.}
    \label{PotenOsc}
\end{figure}

When a scalar field acquires a nonzero vacuum expectation value (VEV), it spontaneously breaks the symmetry of the potential. For definiteness, consider the following potential for a real scalar $\phi$:
\begin{align}
V(\phi) = \frac{1}{2}\,\mu^2(t)\,\phi^2 + \frac{1}{4}\,\lambda\,\phi^4,
\label{HiggsPoten}
\end{align}
where $\mu^2(t)$ oscillates due to an external wave.
When $\mu^2(t) > 0$, the VEV $v_\phi \equiv \langle \phi \rangle$, is $v_\phi = 0$, and the vacuum retains its $\mathbb{Z}_2$ symmetry.
In contrast, if $\mu^2(t) < 0$ (tachyonic), the VEV is $v_\phi = \pm \sqrt{-\mu^2(t) / \lambda}$.
Now, let the sign of $\mu^2(t)$ oscillate over time.
Then, the scalar’s VEV periodically switches between zero and nonzero values, as shown schematically in Fig.~\ref{PotenOsc}.
When $v_\phi \neq 0$, the symmetry of the potential is spontaneously broken.
In contrast, when $v_\phi = 0$, the symmetry is restored.
For simplicity, here we consider a potential with a discrete $\mathbb{Z}_2$ symmetry. However, the same principle applies to continuous symmetries.

Such a time-dependent potential can be realized if $\mu^2(t)$ is generated by an oscillating field, for instance, wave DM.
The wave DM mass ranges from approximately $10^{-22}\,\mathrm{eV}$ (from Lyman-$\alpha$ forest data) up to about $30\,\mathrm{eV}$ (which ensures harmonic oscillation behavior)~\cite{Hui:2021tkt}.
The equation of motion for wave DM $\Phi$ in an expanding universe, assuming spatial homogeneity, is
\begin{align}
\ddot{\Phi} + 3H\,\dot{\Phi} + M_{\Phi}^2\,\Phi = 0,
\label{EoM1}
\end{align}
where $H$ is the Hubble parameter, and $M_\Phi$ is the mass of the wave DM.
In the current epoch, Hubble friction is small compared with $M_\Phi$, so $\Phi(t)$ oscillates as
\begin{align}
\Phi(t) = \frac{\sqrt{2\rho_{\Phi}}}{M_{\Phi}}\,\cos (M_{\Phi} t),
\label{waveDM}
\end{align}
where $\rho_\Phi$ is the wave DM density. 
Many recent works have investigated how particle masses can vary if the coupling of the particle with wave DM directly forms a mass~\cite{VanTilburg:2015oza, Reynoso:2016hjr, Arvanitaki:2016fyj, Berlin:2016woy, Zhao:2017wmo, Krnjaic:2017zlz, Brdar:2017kbt, Davoudiasl:2018hjw, Liao:2018byh, Capozzi:2018bps, Huang:2018cwo, Farzan:2019yvo, Cline:2019seo, Dev:2020kgz, Losada:2021bxx, Huang:2021kam, Chun:2021ief, Dev:2022bae, Huang:2022wmz, Losada:2022uvr, Guo:2022vxr, Brzeminski:2022rkf, Alonso-Alvarez:2023tii, Losada:2023zap, Davoudiasl:2023uiq, Gherghetta:2023myo, ChoeJo:2023ffp, Chen:2023vkq, ChoeJo:2023cnx, Martinez-Mirave:2024dmw, ChoeJo:2024wqr, Plestid:2024kyy, Lee:2024rdc, Lambiase:2025}.
In such scenarios, the mass of a particle changes over time, and it vanishes only momentarily when the wave DM field value equals zero.

In contrast, in our mechanism, it is the \emph{coefficient} $\mu^2$ in $V(\phi)$ that is induced by wave DM, causing the \emph{shape} of the scalar potential to oscillate. 
Consider the scalar potential, which includes a real singlet scalar $\phi$, and a real wave DM scalar $\Phi$.
\begin{align}
V(\phi) = \frac{1}{2}\bigl(\mu_0^2 + \kappa\,\Phi^2\bigr)\,\phi^2
\;+\; \frac{\lambda}{4}\,\phi^4 .
\label{PotentialOsc}
\end{align}
We assume bare term $\mu_0^2 < 0$ (tachyonic) and introduce the scalar mixing coupling $\kappa$ between the singlet scalar $\phi$ and the wave DM $\Phi$.
Both $\kappa$ and $\lambda$ are positive.
Note that $\mu^2(t)$ in Eq.~\eqref{HiggsPoten} is effectively replaced by
\begin{align}
\mu_{\text{eff}}^2(t) \equiv \mu_0^2 + \kappa\,\Phi^2(t),
\label{effmu}
\end{align}
which oscillates over time. 
The sign of the singlet $\phi$ potential’s quadratic term, $\mu_{\text{eff}}^2(t)$, flips if $-\mu_0^2 < \kappa\,\Phi^2$, making $\mu_{\text{eff}}^2(t) > 0$. 
In this case, the symmetry remains unbroken $(v_\phi(t) = 0)$.
Conversely, if $-\mu_0^2 > \kappa\,\Phi^2$ $(\mu_{\text{eff}}^2(t) < 0)$, the symmetry is broken and the singlet acquires a nonzero VEV, $v_\phi(t) = \pm \sqrt{-\mu^2_\mathrm{eff}(t) / \lambda}$.
The effective $\mu^2$ term and thus the singlet scalar potential $V(\phi)$ oscillate with the frequency of the wave DM.

\section{Conditions and constraints for potential oscillation}
\label{ConditionSec}

\begin{figure}[tb]
    \centering
    \subfloat[\label{parameter1}]{%
      \includegraphics[width=.48\textwidth]{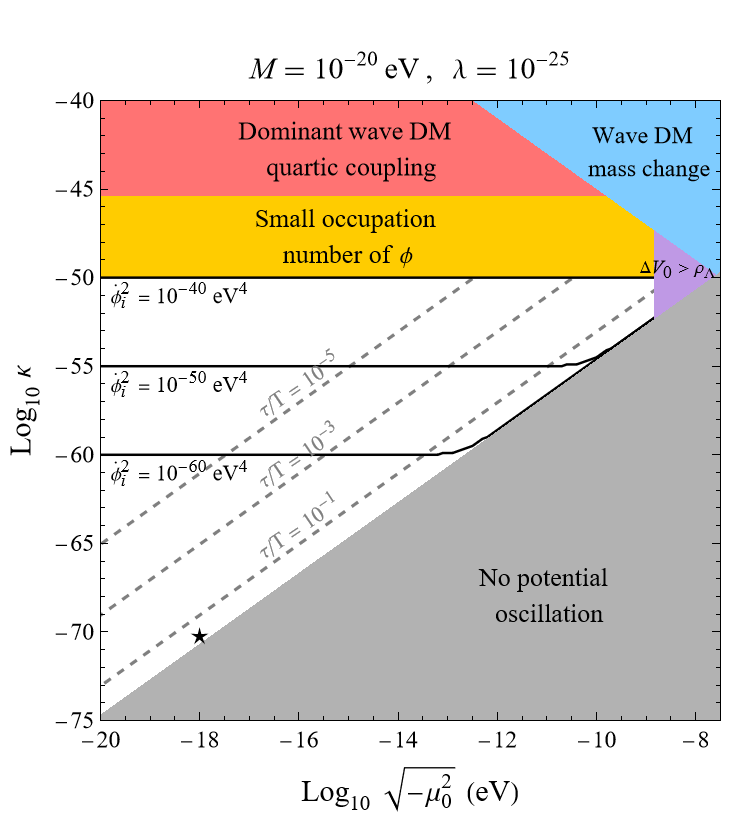}%
    } \quad 
    \subfloat[\label{parameter2}]{%
      \includegraphics[width=.48\textwidth]{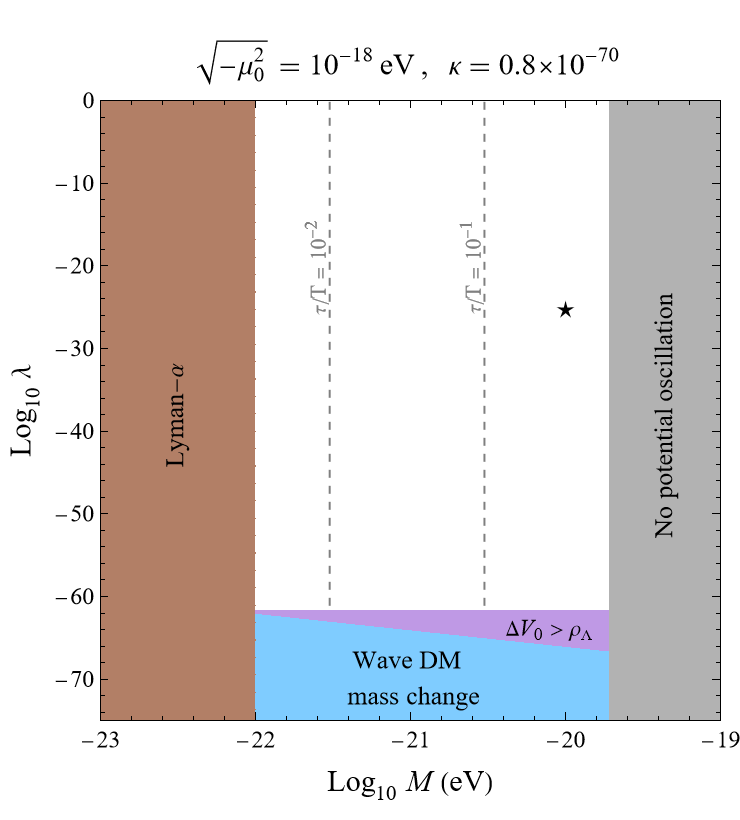}%
    }
    \caption{
    (a) Parameter space in the $(\sqrt{-\mu_0^2}, \kappa)$ plane, where $\mu_0$ is the tachyonic bare mass of the singlet $\phi$ and $\kappa$ is the mixing strength of $\phi$ and wave DM $\Phi$. 
    (b) Parameter space in the $(M, \lambda)$ plane, where $M$ is the bare mass of the wave DM $\Phi$ and $\lambda$ is the singlet $\phi$'s quartic coupling.
    Lines corresponding to the symmetry-broken time ratio $\tau / T$ are shown, where $T \!=\! 2\pi/M_\Phi$.
    The occupation number bound is evaluated at twice the initial kinetic energy of the singlet scalar, $\dot{\phi}_i^{2}=10^{-40}\,\mathrm{eV}^4$. Reducing this initial energy shifts the bound downward.
    The star denotes the benchmark point used in Fig.~\ref{modulation}.
    In (b), the constraint from the dominant quartic coupling of the wave DM is weaker than the Lyman-$\alpha$ bound.
    }
\label{parameter}
\end{figure}

The full Lagrangian with the singlet scalar $\phi$ and the wave DM $\Phi$ is given by
\begin{align}
\mathcal{L} \supset - \frac{1}{2}\bigl(\mu_0^2 + \kappa\,\Phi^2\bigr)\,\phi^2
\;-\; \frac{\lambda}{4}\,\phi^4 
\;-\; \frac{1}{2} M^2 \, \Phi^2 ,
\label{Lagrangian}
\end{align}
where $M$ is the bare mass term of the wave DM.
Figure~\ref{parameter} presents various conditions and constraints on \(\bigl(\sqrt{-\mu_0^2}, \kappa\bigr)\) and \(\bigl(M, \lambda\bigr)\). 

For the potential to oscillate, the wave DM amplitude must exceed the magnitude of the singlet $\phi$’s tachyonic bare mass, thereby allowing $\mu_{\text{eff}}^2(t) > 0$ (gray bound in Fig.~\ref{parameter}):
\begin{align}
-\,\mu_0^2 < \kappa\,\Phi_{\text{max}}^2 \;=\; \frac{2\,\kappa\,\rho_\Phi}{M_\Phi^2}.
\label{Condition1}
\end{align}
When the above condition is satisfied, the time interval $\tau$ for which the symmetry is broken during each period $T = 2\pi / M_\Phi$ is determined by
\begin{align}
\sin^2 \!\Bigl(\frac{\pi}{2}\,\frac{\tau}{T}\Bigr) \;=\; \frac{-\,\mu_0^2}{\,\kappa\,\Phi_{\text{max}}^2\,}.
\label{TimeRatio}
\end{align}
We refer to $\tau/T$ as the \emph{symmetry-broken time ratio.}
If Eq.~\eqref{Condition1} is not satisfied, the potential oscillation does not occur and the symmetry is broken at all times ($\tau/T = 1$).

The effective wave DM mass is given by \(M_\Phi = \sqrt{M^2 + \kappa\,\phi^2}\), but we assume the contribution from the singlet scalar is relatively small so that \(M_\Phi\) remains nearly constant\footnote{In general, one could also consider an oscillating effective mass with sizable changes~\cite{Kofman:1994rk, Kofman:1997yn, Bassett:1997az, Co:2017mop, Nakayama:2021avl}, but here we focus on the case where \(M_\Phi\) remains nearly constant.} by imposing (blue bound in Fig~\ref{parameter}):
\begin{align}
M^2 \gg \kappa \, \phi^2_\mathrm{max} \simeq \frac{-\kappa \mu_0^2}{\lambda}.
\label{Condition2}
\end{align}
Under this condition, the shape and evolution of the wave DM $\Phi$ are nearly unaffected by the singlet scalar $\phi$. In contrast, the oscillating wave DM does affect the potential and evolution of the singlet scalar $\phi$, as already mentioned. By combining Eq.~~\eqref{Condition2} with Eq.~~\eqref{Condition1},
one obtains an upper bound on \(-\mu_0^2\).
\begin{align}
-\,\mu_0^2 < (2\rho_\Phi\,\lambda)^{1/2}.
\label{muBound}
\end{align}
For nearly constant $M_\Phi$ and small $\tau/T$, Eq.~\eqref{TimeRatio} reduces to $\tau \simeq \sqrt{-8\,\mu_0^2 / (\rho_\Phi \,\kappa)}$, independent of $M$.

Although the potential in Eq.~\eqref{Lagrangian} does not include a quartic term for the wave DM $\Phi$, such a term can be generated at the one-loop level via scalar mixing ($\tfrac{1}{2}\,\kappa\,\Phi^2\,\phi^2$).
If this quartic term dominates, the wave DM energy density would behave like radiation~\cite{Turner:1983he, Dev:2022bae}.
Due to the large amplitude of wave DM \(\Phi_\mathrm{max} \propto \sqrt{2 \rho_\Phi} \sim a^{-3/2}\) in the early universe where $a$ is the scale factor, \(\Phi\) may have behaved like dark radiation, necessitating a transition to a matter-like component by at least \(a \sim 10^{-6}\)~\cite{Das:2020nwc, Dev:2022bae}. Consequently, the loop-generated quartic term from the Coleman-Weinberg potential \cite{Coleman:1973jx} at this epoch must be smaller than the quadratic term (red bound in Fig.~\ref{parameter}):
\begin{align}
\frac{\kappa^2}{16\pi^2}\, \left.\Phi_{\mathrm{max}}^4 \right|_{ a=10^{-6}} < \frac{1}{2}\,M^2\, \left.\Phi_{\mathrm{max}}^2 \right|_{ a=10^{-6}} ,
\label{quartic2}
\end{align}
where \( \left. \Phi_{\mathrm{max}} \right|_{ a=10^{-6}} \) is the wave DM amplitude at \(a \sim 10^{-6}\).
Here, \(\rho_\Phi\) is set to be nearly equal to the present-day dark matter density \(\rho_{\mathrm{dm}}^0\).

Similarly, one-loop corrections to the singlet \(\phi\)’s quartic coupling \(\lambda\) may arise. To ensure these corrections do not significantly alter the bare value of \(\lambda\), we impose
\begin{align}
\frac{\kappa^2}{16\pi^2} \;<\; \frac{\lambda}{4}.
\label{quartic1}
\end{align}
Such a constraint from loop-level induced quartic couplings of the singlet scalar \(\phi\) and the wave DM \(\Phi\) requires that the scalar mixing $\kappa$ be feeble, which in turn suppresses the production of \(\phi\) in the early universe~\cite{Hall:2009bx}.
Consequently, its contribution to the dark matter abundance is negligible.\footnote{For the portal term $\kappa \phi^2 \Phi^2$, the singlet scalar $\phi$’s production rate is approximately $\Gamma_\mathrm{\Phi \Phi \rightarrow \phi \phi} \propto \kappa^2 \, T$, where $T$ denotes the temperature of the universe. 
If $\kappa \lesssim 10^{-10}$, then the production rate $\Gamma_{\Phi \Phi \rightarrow \phi \phi}$ remains much smaller than the Hubble expansion rate $H(T)$ throughout cosmic history, so $\phi$ never thermalizes.
Its abundance is then given by $\Omega_\phi\,h^2 \sim 0.1 \left(\tfrac{\kappa}{10^{-12}}\right)^2$ \cite{Hall:2009bx}.
In our scenario, the feeble coupling $\kappa$ prevents $\phi$ from ever thermalizing and from contributing to the dark matter abundance.}
Further cosmological considerations—such as the fast-roll condition and the topological defects arising from symmetry breaking—are presented in Appendices~\ref{cosmologySec} and \ref{DomainWallSec}.

To avoid a large fluctuating vacuum energy, the vacuum energy difference \(\Delta V_0\) between the symmetry-broken and restored phases of the singlet scalar $\phi$'s potential must remain below the energy density associated with the cosmological constant \(\Lambda\). Concretely, we impose the condition (purple bound in Fig.~\ref{parameter})
\begin{align}
\Delta V_0 \;\sim\; \frac{(-\mu_0^2)^2}{\lambda} \;<\; \rho_{\Lambda} \; = \; \frac{\Lambda}{8 \,\pi\, G} \;\sim\; 10^{-47}\,\mathrm{GeV}^4,
\end{align}
where \(\rho_{\Lambda}\) is the vacuum energy density.

\section{Dynamics of the singlet scalar}
\label{DynamicSec}

From the Lagrangian in Eq.~\eqref{Lagrangian}, the coupled equations of motion are given by
\begin{align}
\ddot{\Phi} + 3H\dot{\Phi} + \bigl(M^2 + \kappa\,\phi^2\bigr)\,\Phi = 0,
\label{eqPhi}\\
\ddot{\phi} + 3H\dot{\phi} + \bigl(\mu_0^2 + \kappa\,\Phi^2\bigr)\,\phi + \lambda\,\phi^3 = 0 .
\label{EoM}
\end{align}
To solve the above equations, we impose the following initial conditions. At $t = 0$, the wave DM field is set to $\Phi_{\max}$ and its time derivative to 0. With this choice, the singlet scalar $\phi$ lies in the symmetry-restored phase at $t = 0$, so its initial field value is set to be 0. The only remaining free parameter is the singlet scalar’s initial velocity $\dot{\phi}_{i}$ or, equivalently, its initial kinetic energy $\tfrac12 \dot{\phi}_{i}^{\,2}$.

The energy density $\rho_\phi(t)$ of the singlet scalar is
\begin{align}
\rho_\phi(t) & \equiv \frac{1}{2} \dot{\phi}(t)^2 + V(\phi) ,
\label{phiErg}
\end{align}
where the first term represents the kinetic energy of the singlet scalar.
The singlet scalar $\phi$ oscillates around its VEV $v_\phi$.
The singlet's oscillation amplitude $\phi_{\text{amp}}(t)$ from its VEV satisfies,
\begin{align}
\rho_\phi(t) = V\Big( v_\phi(t) + \phi_{\text{amp}}(t) \Big) ,
\label{phiamp}
\end{align}
At this moment, the kinetic energy of the singlet scalar vanishes.
By solving Eq.~\eqref{phiamp}, one can determine the oscillation amplitude $\phi_\mathrm{amp}$ around the singlet scalar’s oscillation center $v_\phi$. In the symmetry-restored phase, the amplitude $\phi_\mathrm{amp}^\mathrm{restored}$ must remain small compared to the nonzero VEV $v_\phi$. Otherwise, it becomes difficult to distinguish the scale of the singlet scalar field $\phi$ in the symmetry-restored and symmetry-breaking phases. We will address this further by numerically solving the equations of motion.
The mass $m_\phi(t)$ of the singlet is given by the quadratic term in the expansion of the singlet potential at the VEV.
\begin{align}
m_\phi(t)
= \left. \frac{\partial^2 V}{\partial \phi^2} \right|_{\phi = v_\phi}
= \begin{cases}
\sqrt{-2 \mu_{\text{eff}}^2(t)} & \big(-\mu_0^2 > \kappa \Phi^2(t) \big) 
\\
\mu_{\text{eff}}(t) & \big(-\mu_0^2 < \kappa \Phi^2(t)\big) .
\end{cases}
\label{phimass}
\end{align}
The singlet mass $m_\phi$ vanishes when the sign of $\mu_\text{eff}^2$ is changed.
The oscillation frequency $f_\phi(t)$ of the singlet is
\begin{align}
\frac{1}{f_\phi(t)} = \frac{4}{\sqrt{m_\phi^2(t) + \lambda \phi_{\text{amp}}^2(t)}} K\left( \sqrt{\frac{\lambda \phi_{\text{amp}}^2(t)}{2 m_\phi^2(t) + \lambda \phi_{\text{amp}}^2(t)}} \right) ,
\label{phifreq}
\end{align}
where $K$ is the complete elliptic integral of the first kind arsing from non-linear effect of the $\phi^3$ term in Eq.~\eqref{EoM}.
Due to the non-linear effect ($\lambda \neq 0$), the frequency $f_\phi$ does not vanish.
In the $\lambda \rightarrow 0$ limit, the frequency in Eq.~\eqref{phifreq} reduces to $f_\phi = m_\phi / 2\pi$ with $K(0) = \pi/2$.

To treat the singlet scalar $\phi$ as a classical field, its occupation number $N_\phi$ within a de Broglie volume must greatly exceed 1 (yellow bound in Fig.~\ref{parameter}) so that quantum fluctuations are negligible \cite{Hui:2021tkt}:
\begin{align}
N_\phi(t)=\frac{\lvert\rho_\phi(t)\rvert}{m_\phi^{4}(t)\,v_{\text{gal}}^{3}}\gg1,
\label{Nphi}
\end{align}
where $v_{\text{gal}}=250~\text{km/s}$ is the characteristic velocity dispersion of the Galactic halo.
This bound depends on twice the initial kinetic energy $\dot{\phi}_i^{\,2}$ of the singlet scalar at $t=0$; lowering that energy shifts the constraint downward.
Large values of $\kappa$ are excluded because the singlet scalar mass $m_\phi$ becomes significant in the symmetry-restored phase, temporarily driving $N_\phi(t)$ below 1.

\begin{figure*}[tb]
    \centering
    \subfloat[\label{parameter1}]{%
      \includegraphics[width=.48\linewidth]{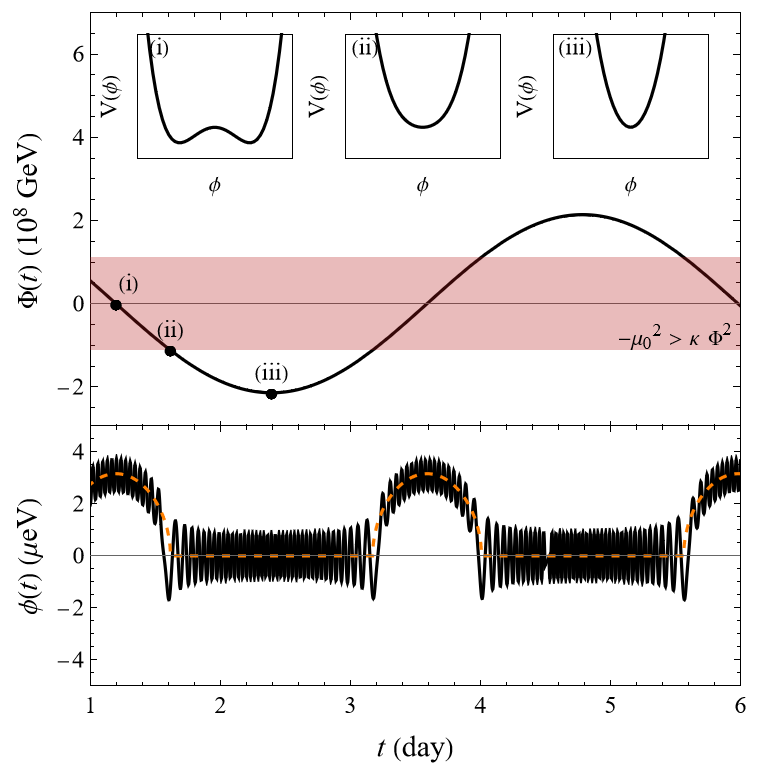}%
    } \quad 
    \subfloat[\label{parameter2}]{%
      \includegraphics[width=.48\linewidth]{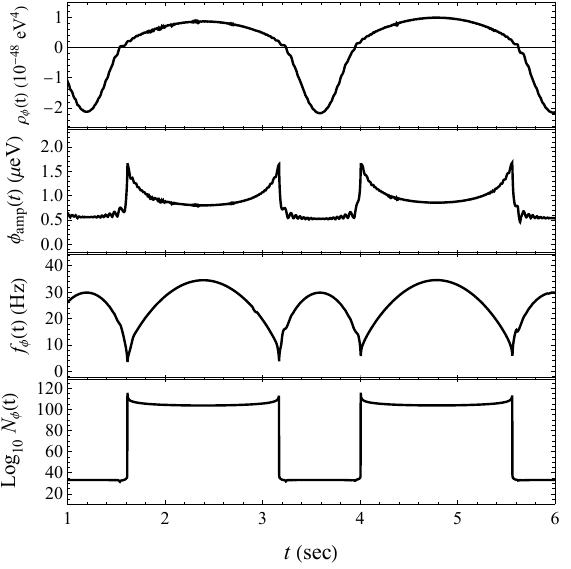}%
    }
    \caption{
    (a) Time evolution of the wave DM field \(\Phi(t)\), the singlet scalar potential \(V(\phi)\), and hence the singlet scalar field \(\phi(t)\). The benchmark parameters are \(M = 10^{-20}\,\mathrm{eV}\), \(\lambda = 10^{-25}\), \(\sqrt{-\mu_0^2} = 10^{-18}\,\mathrm{eV}\), and \(\kappa = 0.8 \times 10^{-70}\).
    At $t = 0$, twice the initial kinetic energy of $\phi$ is set to $\dot{\phi}_i^{\,2} \simeq 10^{-51}\,\mathrm{eV}^4$, a value that is significantly small compared with the wave DM energy density. Because of the suppressed production of the singlet scalar $\phi$, its energy density can also remain naturally small.
    Inside the red bands, the potential is tachyonic \(\bigl(\mu_\text{eff}^2(t) < 0\bigr)\), causing the symmetry to break. Outside those bands, the symmetry is restored.
    The period is \( T = 2\pi / M_\Phi \simeq 4.8\,\mathrm{days}  \) and the symmetry-broken time ratio is $\tau / T \simeq 0.35$.
    The dashed orange curve in the bottom panel represents the VEV $|v_\phi|$ of the singlet scalar.
    (b) The energy density $\rho_\phi(t)$, the oscillation amplitude $\phi_\text{amp}(t)$, the oscillation frequency $f_\phi(t)$, and the occupation number $N_\phi(t)$ of the singlet scalar field $\phi$.
    Some kink structures arise from numerical issues.
    }
    \label{modulation}
\end{figure*}

In Fig.~\ref{modulation}, we illustrate the evolution of the wave DM $\Phi(t)$ and the singlet scalar \(\phi(t)\) for a benchmark point that satisfies all conditions and is consistent with the relevant constraints.
Here we set the scale of the bare mass terms $M$ and $\sqrt{-\mu_0^2}$ are not too far to show both oscillations of $\Phi$ and $\phi$.
The coupling $\kappa$ is properly chosen for a comparable time ratio $\tau/T$ according to Eq.~\eqref{TimeRatio}.
The benchmark yields a period of several days \(\bigl(T \simeq 4.8\,\mathrm{days}\bigr)\) and a comparable time ratio between the symmetry-broken and restored phases \(\bigl(\tau / T \simeq 0.35\bigr)\). Here, we assume that, at present, the wave DM \(\Phi\) constitutes nearly all of the local dark matter density \(\rho_\mathrm{dm}^0 = 0.3\,\mathrm{GeV}/\mathrm{cm}^3\). 
The nonzero VEV of the singlet scalar \(\phi\) slightly increases \(M_\Phi\) (via \(\kappa\,\phi^2\)), thereby speeding up oscillations and shortening the period during which \(\mu_\text{eff}^2(t) < 0\).

The time evolution of the energy density \(\rho_\phi(t)\), the oscillation amplitude \(\phi_\text{amp}(t)\), and the oscillation frequency \(f_\phi(t)\) of the singlet scalar \(\phi\) are shown together in Fig.~\ref{modulation}. Although energy is transferred between the singlet scalar and wave DM, this transfer is extremely small compared to the wave DM density (\(|\Delta \rho_\phi| \ll \rho_\Phi\)). In the symmetry-broken phase, the singlet loses energy \((\rho_\phi < 0)\). In contrast, in the symmetry-restored phase, the singlet gains energy \((\rho_\phi > 0)\). One can show that, by solving Eq.~\eqref{phiamp} for $\phi_\mathrm{amp}$, the oscillation amplitude \(\phi_\text{amp}\) is larger in the symmetry-restored phase than in the symmetry-broken phase \((\phi_\text{amp}^\text{restored} > \phi_\text{amp}^\text{broken})\), yet it remains below the maximum of the nonzero VEV \((\phi_\text{amp}^\text{restored} < |v_\phi^\text{max}|)\).
In Fig.~\ref{modulation}, we set twice the initial kinetic energy of the singlet scalar to $\dot{\phi}_{i}^{\,2} \simeq 10^{-51}\,\mathrm{eV}^{4}$. With this choice, the singlet scalar’s occupation number $N_\phi(t)$ remains greater than unity throughout the evolution.

\section{Implications for neutrino physics}
\label{CnuBsec}

As a concrete application of potential oscillation, we examine the neutrino sector.
Let the Majorana mass term of the sterile neutrino \(N\) be generated by the singlet scalar \(\phi\) associated with the potential oscillation.
\begin{align}
\mathcal{L}_\mathrm{mass} &= - \frac{y}{2}\,\phi\,\overline{N^c}\,N \;+\; \text{h.c.} \label{Majorana}
\end{align}
The Majorana mass is \(M_N = y\,\phi\).
When the symmetry of the singlet potential is broken, the Majorana mass $M_N$ oscillates around a nonzero VEV $v_\phi$.
In contrast, when the symmetry is restored, $M_N$ oscillates at 0.

Also, the radiatively generated quartic coupling for the singlet \(\phi\) arising from this Yukawa interaction should not significantly alter the bare term.
\begin{align}
\frac{y^4}{16\pi^2} \;<\; \frac{\lambda}{4} .
\label{quartic3}
\end{align}
Combining Eq.~\eqref{quartic3} with Eq.~\eqref{muBound}, the upper bound on the Majorana mass, \(M_N < y \sqrt{-\mu_0^2 / \lambda}\), becomes
\begin{align}
M_N < (8\,\pi^2\,\rho_\Phi)^{1/4}\,\simeq\,0.12\,\mathrm{eV}.
\label{MNbound}
\end{align}
Because $M_N$ is bounded, the Dirac mass $m_D$ must also be very small to account for the observed small active neutrino mass.
In addition, the coupling \(y\) may also be constrained by majoron-emitting decays \cite{KamLAND-Zen:2012uen}, although these bounds can vary depending on the chosen value of \(\lambda\).

\begin{figure}[tb]
    \centering
    \includegraphics[width=0.65\textwidth]{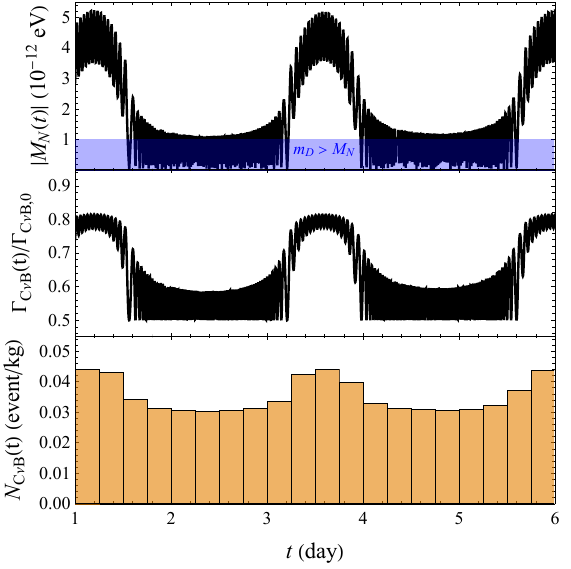}
    \caption{The modulation of the Majorana mass term $|M_N(t)|$ and the C$\nu$B detection rate $\Gamma_{\text{C}\nu\text{B}}(t)$ over the reference value $\Gamma_{\text{C}\nu\text{B},0}$ for no active-sterile neutrino mixing ($\sin^2 \theta_{as} = 0$).
    The parameter setup is the same as Fig.~\ref{modulation}.
    The Majorana mass $M_N$ touches zero, but it is not shown properly due to the resolution issue.
    The value of $y$ is taken by the possible maximum according to Eq.~\eqref{quartic3}.
    The Dirac mass is fixed at $m_D = 10^{-12}\,\mathrm{eV}$, providing the lightest neutrino mass scale as $\sim 10^{-13}\,\mathrm{eV}$, and the effect of three active neutrino generations is included with values taken from Ref.~\cite{ParticleDataGroup:2024cfk}.
    For the blue band, the neutrinos are in the quasi-Dirac state ($m_D > M_N$).
    In this region, the active-sterile mixing angle is nearly maximal, so the C$\nu$B detection rate becomes small.
    When the mixing is maximal, the ratio becomes $\Gamma_{\text{C}\nu\text{B}}(t)/\Gamma_{\text{C}\nu\text{B},0} = 0.5$.
    Since the maximum of $M_N$ is not much farther from $m_D$, the ratio does not touch 1.
    The lowest panel shows the cosmic neutrino-capture event rate $N_{\text{C}\nu\text{B}}(t)$, normalized to the tritium target mass $M_T$, for our benchmark point, using a 6-hour time bin $\Delta t_\mathrm{bin}$.
    The PTOLEMY design envisions a 100 g tritium target~\cite{Long:2014zva}.
    }
\label{CnuBplot}
\end{figure}

Depending on the shape of \(\phi\)’s potential at a given time, neutrinos can alternate effective type between a quasi-Dirac state, characterized by a small ratio of Majorana to Dirac mass \((M_N \ll m_D)\), and a Majorana state \((M_N \gg m_D)\). 
As shown in Fig.~\ref{CnuBplot}, one can set the Dirac mass $m_D$ properly so that neutrinos cross over two regions.
Consequently, this leads to distinctive predictions, including periodic modulation signals in the cosmic neutrino.

The C$\nu$B consists of relic neutrinos that decoupled in the early universe, now cooled to $T_\nu \simeq 0.168\,\mathrm{meV}$ and in a non-relativistic state. Experiments such as PTOLEMY~\cite{Betts:2013uya, Long:2014zva, PTOLEMY:2018jst, PTOLEMY:2019hkd} aim to detect the C$\nu$B through neutrino capture on tritium ($\nu_e + {}^3\text{H} \to {}^3\text{He} + e^-$).
It is a well-known fact that the capture rate for Majorana neutrinos is twice that of Dirac neutrinos, $\Gamma_{\text{C}\nu\text{B}}^M = 2\,\Gamma_{\text{C}\nu\text{B}}^D$~\cite{Long:2014zva}.

This can be generalized to include the quasi-Dirac limit of the neutrino~\cite{Perez-Gonzalez:2023llw}. 
For the sake of simplicity, we discuss this with a single-flavor active–sterile mixing scenario.
By including the effect of active-sterile neutrino mixing, the C\(\nu\)B detection rate becomes 
$\Gamma_{\text{C}\nu\text{B}} \propto \bigl(1 - P(\nu_a \rightarrow \nu_s)\bigr)$ since only the active cosmic neutrino component can be detected.
In the seesaw limit \((M_N \gg m_D)\), the active-sterile mixing is extremely small, with \(\sin^2 \theta_{as} \simeq (m_D / M_N)^2 \rightarrow 0\). Consequently, the mixing probability from an active neutrino \(\nu_a\) to a sterile neutrino \(\nu_s\) vanishes:
$P(\nu_a \rightarrow \nu_s) \sim \tfrac{1}{2}\,\sin^2(2\theta_{as}) \;\rightarrow\; 0.$\footnote{The phase term of the mixing probability can be averaged out since neutrinos propagate over cosmic timescales after decoupling~\cite{Perez-Gonzalez:2023llw}.}

In contrast, in the quasi-Dirac case \((M_N \ll m_D)\), the mass eigenvalues are nearly degenerate at the Dirac mass \(m_D\) with a small splitting, and the active-sterile neutrino mixing is nearly maximal \(\bigl(\sin^2 \theta_{as} \simeq \tfrac{1}{2}\bigr)\). Hence, the mixing probability approaches 
$P(\nu_a \rightarrow \nu_s) \;\rightarrow\; \tfrac{1}{2}.$
The C\(\nu\)B detection rate \(\Gamma_{\text{C}\nu\text{B}}\) is comparatively smaller in the quasi-Dirac limit.

Figure~\ref{CnuBplot} illustrates how the detection rate is modulated by neutrino oscillations between quasi-Dirac and Majorana states. The \(\sim 12.3\)-year half-life of tritium \(\beta\)-decay does not affect short-timescale modulation analyses. Moreover, the \(\beta\)-decay background can be distinguished and subtracted based on its spectral characteristics relative to the C\(\nu\)B capture signal.
Due to its unique signatures, a modulation-based search may offer greater advantages than a straightforward event-counting approach by effectively canceling out systematic uncertainties. A detailed sensitivity analysis of the C$\nu$B modulation search is presented in Appendix~\ref{ModulationSec}.

A time-dependent Majorana mass term can affect active neutrino masses and flavor oscillations. The eigenvalues of both active and sterile neutrinos depend on the Dirac mass \(m_D\) and the Majorana mass \(M_N\).
If the Majorana mass is small (as shown in Eq.~\eqref{MNbound}), the variations in the active neutrino mass and the mass square differences remain naturally modest.
Consequently, their impact on neutrino flavor oscillations is negligible.

Before neutrino decoupling, the production rate of the sterile neutrino via the mixing is $\Gamma_{a\rightarrow s}(T) \simeq \frac{1}{2} \sin^2 2 \theta_{as}^m(T) \Gamma_\text{int}(T)$~\cite{Kainulainen:1990ds, Enqvist:1991qj, Hannestad:2012ky}, where $T$ is the temperature of the universe, $\sin^2 2 \theta_{as}^m$ is active-sterile matter mixing in the $\ell$, $\nu$ plasma, and $\Gamma_\text{int} \simeq G_F^2 T^5$ is the scattering rate of the active neutrino.
For a tiny mass splitting $\Delta m_{as}^2$ between active and sterile states, the mixing is suppressed by $\sin^2 2 \theta_{as}^m \simeq \big[\Delta m_{as}^2 / (G_F T^4) \big]^2$.
In our scenario with $\Delta m_{as}^2 \sim 10^{-24}\,\mathrm{eV^2} \ll G_F T^4$, the production rate is strongly suppressed ($\Gamma_{a\rightarrow s} \ll H$) preventing sterile neutrinos from ever thermalizing.
Because the sterile neutrino density in today’s universe is negligible, one can neglect the additional term in Eq.~\eqref{EoM} involving sterile neutrinos $N$ induced from the Yukawa interaction in Eq.~\eqref{Majorana}.
Moreover, because sterile neutrinos are produced so rarely, the Yukawa interaction makes a negligible contribution to the singlet scalar $\phi$’s abundance.
Cosmological and laboratory data impose no additional constraints on such a small sterile neutrino mass scale~\cite{Vincent:2014rja, Bolton:2019pcu}.

\section{Summary and outlook}
\label{SummarySec}

We have introduced a potential oscillation mechanism that periodically breaks and restores symmetry through an externally modulated scalar potential. Symmetry and its breaking lie at the heart of many areas of physics. While our main application focuses on neutrino physics, the mechanism naturally extends to a wide range of frameworks with suitable symmetries. It provides an elegant way to realize time-dependent mass structures and offers rich phenomenological signatures, such as modulated cosmic neutrino background detection rates.

\vspace{2mm}
\begin{acknowledgments}
We thank Jiheon Lee and Jaeok Yi for their helpful comments.
This work was supported in part by the National Research Foundation of Korea (Grant No. RS-2024-00352537).
\end{acknowledgments}

\appendix

\section{Implications for cosmology}
\label{cosmologySec}

In this section, we discuss the evolution of the wave DM amplitude over cosmic time, as well as the general conditions under which the singlet scalar $\phi$ performs fast rolls.

In the case of the wave DM, oscillations start when $H \simeq M_\Phi$.
Over cosmic history, the wave DM amplitude \(\Phi_{\text{max}}\) evolved with the universe expansion.
Since \(\Phi_{\text{max}} \propto \sqrt{2\,\rho_\Phi} \sim a^{-3/2}\), with the scale factor $a$, it was much larger at earlier times. As a result, \(-\mu_0^2 \ll \kappa\,\Phi_{\text{max}}^2\). Consequently, from Eq.~\eqref{TimeRatio}, the symmetry-broken phase persisted only for an extremely short time, \(\tau/T \ll 1\).
This short symmetry-broken interval implies that, for most of the time, the potential remained in the symmetry-restored phase and was dominated by thermal effects, similar to standard early universe phase transitions.

For the singlet $\phi$ field to track the minimum of its potential, it must roll quickly (i.e., avoid the slow-roll regime). For the singlet scalar potential $V(\phi)$ from Eq.~\eqref{PotentialOsc}, this fast-roll condition can be written as
\begin{align}
H^2 \;\ll\;  \frac{\partial^2 V(\phi)}{\partial \phi^2} 
\;=\;  (\mu_0^2 + \kappa\,\Phi^2) \;+\; \lambda\,\phi^2,
\label{NotSlow}
\end{align}
so that the Hubble friction term is negligible in the equation of motion.
Since the current Hubble parameter value is $H_0 \simeq 10^{-33}\,\mathrm{eV}$, this condition is easily satisfied in the present epoch with appropriate parameter choices.
As an example, using the benchmark parameters in Fig.~\ref{modulation}, the right-hand side of condition~\eqref{NotSlow} evaluates to $\sim 10^{-36}\,\mathrm{eV}^2 \gg H_0^2$, fulfilling the fast-roll requirement.

However, in the early universe, the large Hubble parameter prevented the singlet scalar \(\phi\) from satisfying the fast-roll condition in Eq.~\eqref{NotSlow}. The scale factor \(a_\ast\), at which the singlet begins its fast-roll, is approximately given by  
\begin{align}
\frac{a_\ast}{a_{\mathrm{eq}}} \simeq
\begin{cases}
\displaystyle
\left( \frac{H_{\mathrm{eq}}^{7/2} \, M^{1/2}}{2 \, \kappa \, \rho_{\mathrm{dm}}^\mathrm{eq}} \right)^{1/4} 
& \text{if } H_{\mathrm{eq}}^{3/2} \, M^{5/2} < 2 \, \kappa \, \rho_{\mathrm{dm}}^\mathrm{eq}, \\[2em]
\displaystyle
\frac{H_{\mathrm{eq}}^2 \, M^2}{2 \, \kappa \, \rho_{\mathrm{dm}}^\mathrm{eq}} 
& \text{if } H_{\mathrm{eq}}^{3/2} \, M^{5/2} > 2 \, \kappa \, \rho_{\mathrm{dm}}^\mathrm{eq},
\end{cases}
\end{align}
where \(a_{\mathrm{eq}}\), \(H_{\mathrm{eq}}\), and \(\rho_{\mathrm{dm}}^\mathrm{eq}\) denote the scale factor, the Hubble parameter, and the dark matter density at matter-radiation equality, respectively.
In the first case, \(a_\ast\) precedes the onset of wave DM $\Phi$ oscillations, whereas in the second case, \(a_\ast\) occurs later. In our benchmark parameter setup (see Fig.~\ref{modulation}), fast-roll begins at \(a_\ast \sim 10^{-13}\), while the wave DM field $\Phi$ starts oscillating at \(a \sim 10^{-7}\).

\section{Domain walls}
\label{DomainWallSec}

The breaking of a discrete symmetry can produce domain walls as topological defects~\cite{Zeldovich:1974uw}.
To avoid a cosmological problem, the domain wall energy within the current horizon, \(\sigma_{\mathrm{dw}}\, H_0^{-2}\), must be much smaller than the total energy of the present horizon, \(\rho_c\, H_0^{-3}\). Here, \(\sigma_{\mathrm{dw}} = \tfrac{2\sqrt{2}}{3}\,\lambda^{1/2}\,v_\phi^3\) is the surface energy density of the domain walls, and \(\rho_c\) is the critical energy density of the universe. This requirement simplifies to $
\sigma_{\mathrm{dw}} \ll \rho_c \, H_0^{-1} \,\sim\, (10\,\mathrm{MeV})^3$.

A stronger bound arises from domain wall formation in the early universe.
Domain walls can act as an additional source of anisotropy in the cosmic microwave background (CMB). For consistency with observations, we therefore require \(\sigma_{\mathrm{dw}} \lesssim (1\,\mathrm{MeV})^3\)~\cite{Zeldovich:1974uw, Sikivie:1982qv, Vilenkin:1984ib, Lazanu:2015fua}.

In our scenario, when the singlet \(\phi\) acquires a nonzero VEV \(\bigl(\mu_{\text{eff}}^2(t) = \mu_0^2 + \kappa \Phi^2(t) < 0\bigr)\), the surface energy density is bounded above by
\begin{align}
\sigma_{\mathrm{dw}} 
= \frac{2\sqrt{2}}{3} \,\frac{\bigl(-\mu_0^2 - \kappa \Phi^2\bigr)^{3/2}}{\lambda}
< \frac{2\sqrt{2}}{3} \,\frac{\bigl(2\,\rho_\Phi\bigr)^{3/4}}{\lambda^{1/4}},
\end{align}
at all times in cosmic history, as implied by Eq.~\eqref{muBound}.
Furthermore, the maximum possible value of \(\sigma_{\mathrm{dw}}\) remains below the cosmological bound as long as \(\lambda \gtrsim 10^{-88}\).

Because of this upper bound, domain-wall constraints are weaker than those shown in Fig.~\ref{parameter}. Adopting the values of \(\mu_0\) and \(\lambda\) from the benchmark scenario in Fig.~\ref{modulation}, the surface energy density is estimated to be \(\sigma_{\mathrm{dw}} \sim 10^{-29}\,\mathrm{eV}^3\), which is extremely small.

The dynamics of domain walls under the repetition of symmetry restoration and breaking may differ from the standard scenario. Depending on the oscillation frequency, domain walls may fail to form or be annihilated, potentially relaxing the aforementioned constraints even further.
However, we do not analyze these effects in detail here.

Lastly, domain wall fluctuations can evolve alongside density perturbations, remaining compatible with observations as long as the domain wall’s energy density remains small, consistent with CMB bounds~\cite{Fabris:2000qz, Friedland:2002qs}.
Moreover, gravitational waves may be produced from domain wall annihilation or perturbation~\cite{Vilenkin:1981zs, Hiramatsu:2013qaa, Saikawa:2017hiv, Kitajima:2023cek, Lu:2024dzj}, providing a potential observational signature for the proposed potential oscillation scenario.

\section{A modulation search on C$\nu$B}
\label{ModulationSec}

In this section, we discuss the advantages of modulation-based searches over simple event-counting searches.

One can fit the C$\nu$B count rate in the form
\begin{equation}
N_{\text{C}\nu\text{B}}(t_i) \;\simeq\; N_0 \,\Bigl[\,1 \;+\; \delta \,\cos\bigl(M_\Phi \,t_i\bigr)\Bigr],
\end{equation}
where $N_0$ and $\delta$ represent the modulation center and amplitude fraction, respectively. Here, $t_i = t_0 + i \,\Delta t_\mathrm{bin}$ ($i = 0, 1, 2, \dots$) with bin time $\Delta t_\mathrm{bin}$.

The modulation fraction $\delta$ is approximately
\begin{equation}
\delta \;\simeq\; \frac{\bigl|P_{\max}(\nu_a \to \nu_a) \;-\; P_{\min}(\nu_a \to \nu_a)\bigr|}{\,P_{\max}(\nu_a \to \nu_a) \;+\; P_{\min}(\nu_a \to \nu_a)\,},
\end{equation}
where $P_{\max}(\nu_a \to \nu_a)$ and $P_{\min}(\nu_a \to \nu_a)$ are the maximum and minimum survival probabilities over one period, respectively.

For a simple event-counting measurement with a flat signal, the signal-to-noise ratio (SNR) is
\begin{equation}
\mathrm{SNR}_\mathrm{flat} \;=\; 
\frac{N_0}{\sqrt{ \sigma_\mathrm{stat}^2 + \sigma_\mathrm{sys}^2}},
\label{flatSNR}
\end{equation}
where $\sigma_\mathrm{stat} = \frac{(N_0 + N_B)}{T_\mathrm{exp}}$ is the statistical uncertainty from the Poisson distribution, and $\sigma_\mathrm{sys}$ is the time-independent systematic uncertainty. Here, $N_B$ is the background count rate, and $T_\mathrm{exp}$ is the exposure time for collecting the C$\nu$B signal.

In a modulation analysis, one fits only the oscillatory component, so systematic uncertainties largely cancel out.
Because of the absence of systematic uncertainty, such a modulation-based analysis could achieve higher sensitivity than a straightforward event-counting approach. The corresponding SNR is roughly
\begin{equation}
\mathrm{SNR}_\mathrm{modul} \;=\; \frac{N_0 \,\delta}{\sqrt{\frac{(N_0 + N_B)}{T_\mathrm{exp} /2}}}
\;\simeq\; 4.3\,\delta \,\sqrt{\frac{M_T}{1\,\mathrm{kg}} \,\frac{T_\mathrm{exp}}{1\,\mathrm{yr}}},
\label{modulSNR}
\end{equation}
where $M_T$ is the target mass, and the factor of $1/2$ arises from the orthogonality of the harmonic functions. The background count rate (e.g., from $\beta$-decay tails or external sources) in the energy bin $\Delta_\mathrm{res} \sim 0.1\,\mathrm{eV}$ \cite{Betts:2013uya,Long:2014zva} is taken as
\begin{equation}
N_B \;\simeq\; 0.3\, \frac{M_T}{1\,\mathrm{kg}} \,\frac{\Delta t_\mathrm{bin}}{1\,\mathrm{yr}}.
\end{equation}
Using the benchmark from Fig.~\ref{CnuBplot}, we find $\delta \simeq \mathcal{O}(0.1)$, which requires $M_T \, T_\mathrm{exp} \gtrsim \mathcal{O}(1)\,\mathrm{kg}\cdot\mathrm{yr}$ for $\mathrm{SNR}_\mathrm{modul} \gtrsim 1$.

We note that there have been proposals to search for the C$\nu$B using solar gravitational focusing \cite{Safdi:2014rza}. This method predicts a relatively small modulation amplitude, $\delta \sim \mathcal{O}(0.001 - 0.01)$, which is significantly smaller than the amplitude fractions of $\mathcal{O}(0.1)$ considered in our scenario.

\bibliographystyle{JHEP}
\bibliography{ref.bib}

\providecommand{\href}[2]{#2}\begingroup\raggedright\begin{thebibliography}{10}

\bibitem{Englert:1964et}
F.~Englert and R.~Brout, \emph{{Broken Symmetry and the Mass of Gauge Vector
  Mesons}}, \href{https://doi.org/10.1103/PhysRevLett.13.321}{\emph{Phys. Rev.
  Lett.} {\bfseries 13} (1964) 321}.

\bibitem{Higgs:1964pj}
P.W.~Higgs, \emph{{Broken Symmetries and the Masses of Gauge Bosons}},
  \href{https://doi.org/10.1103/PhysRevLett.13.508}{\emph{Phys. Rev. Lett.}
  {\bfseries 13} (1964) 508}.

\bibitem{Guralnik:1964eu}
G.S.~Guralnik, C.R.~Hagen and T.W.B.~Kibble, \emph{{Global Conservation Laws
  and Massless Particles}},
  \href{https://doi.org/10.1103/PhysRevLett.13.585}{\emph{Phys. Rev. Lett.}
  {\bfseries 13} (1964) 585}.

\bibitem{Kofman:1994rk}
L.~Kofman, A.D.~Linde and A.A.~Starobinsky, \emph{{Reheating after inflation}},
  \href{https://doi.org/10.1103/PhysRevLett.73.3195}{\emph{Phys. Rev. Lett.}
  {\bfseries 73} (1994) 3195}
  [\href{https://arxiv.org/abs/hep-th/9405187}{{\ttfamily hep-th/9405187}}].

\bibitem{Kofman:1997yn}
L.~Kofman, A.D.~Linde and A.A.~Starobinsky, \emph{{Towards the theory of
  reheating after inflation}},
  \href{https://doi.org/10.1103/PhysRevD.56.3258}{\emph{Phys. Rev. D}
  {\bfseries 56} (1997) 3258}
  [\href{https://arxiv.org/abs/hep-ph/9704452}{{\ttfamily hep-ph/9704452}}].

\bibitem{Bassett:1997az}
B.A.~Bassett and S.~Liberati, \emph{{Geometric reheating after inflation}},
  \href{https://doi.org/10.1103/PhysRevD.60.049902}{\emph{Phys. Rev. D}
  {\bfseries 58} (1998) 021302}
  [\href{https://arxiv.org/abs/hep-ph/9709417}{{\ttfamily hep-ph/9709417}}].

\bibitem{Co:2017mop}
R.T.~Co, L.J.~Hall and K.~Harigaya, \emph{{QCD Axion Dark Matter with a Small
  Decay Constant}},
  \href{https://doi.org/10.1103/PhysRevLett.120.211602}{\emph{Phys. Rev. Lett.}
  {\bfseries 120} (2018) 211602}
  [\href{https://arxiv.org/abs/1711.10486}{{\ttfamily 1711.10486}}].

\bibitem{Nakayama:2021avl}
K.~Nakayama and W.~Yin, \emph{{Hidden photon and axion dark matter from
  symmetry breaking}},
  \href{https://doi.org/10.1007/JHEP10(2021)026}{\emph{JHEP} {\bfseries 10}
  (2021) 026} [\href{https://arxiv.org/abs/2105.14549}{{\ttfamily
  2105.14549}}].

\bibitem{Kirzhnits:1972ut}
D.A.~Kirzhnits and A.D.~Linde, \emph{{Macroscopic Consequences of the Weinberg
  Model}}, \href{https://doi.org/10.1016/0370-2693(72)90109-8}{\emph{Phys.
  Lett. B} {\bfseries 42} (1972) 471}.

\bibitem{Senaha:2020mop}
E.~Senaha, \emph{{Symmetry Restoration and Breaking at Finite Temperature: An
  Introductory Review}},
  \href{https://doi.org/10.3390/sym12050733}{\emph{Symmetry} {\bfseries 12}
  (2020) 733}.

\bibitem{Hui:2021tkt}
L.~Hui, \emph{{Wave Dark Matter}},
  \href{https://doi.org/10.1146/annurev-astro-120920-010024}{\emph{Ann. Rev.
  Astron. Astrophys.} {\bfseries 59} (2021) 247}
  [\href{https://arxiv.org/abs/2101.11735}{{\ttfamily 2101.11735}}].

\bibitem{VanTilburg:2015oza}
K.~Van~Tilburg, N.~Leefer, L.~Bougas and D.~Budker, \emph{{Search for
  ultralight scalar dark matter with atomic spectroscopy}},
  \href{https://doi.org/10.1103/PhysRevLett.115.011802}{\emph{Phys. Rev. Lett.}
  {\bfseries 115} (2015) 011802}
  [\href{https://arxiv.org/abs/1503.06886}{{\ttfamily 1503.06886}}].

\bibitem{Reynoso:2016hjr}
M.M.~Reynoso and O.A.~Sampayo, \emph{{Propagation of high-energy neutrinos in a
  background of ultralight scalar dark matter}},
  \href{https://doi.org/10.1016/j.astropartphys.2016.05.004}{\emph{Astropart.
  Phys.} {\bfseries 82} (2016) 10}
  [\href{https://arxiv.org/abs/1605.09671}{{\ttfamily 1605.09671}}].

\bibitem{Arvanitaki:2016fyj}
A.~Arvanitaki, P.W.~Graham, J.M.~Hogan, S.~Rajendran and K.~Van~Tilburg,
  \emph{{Search for light scalar dark matter with atomic gravitational wave
  detectors}}, \href{https://doi.org/10.1103/PhysRevD.97.075020}{\emph{Phys.
  Rev. D} {\bfseries 97} (2018) 075020}
  [\href{https://arxiv.org/abs/1606.04541}{{\ttfamily 1606.04541}}].

\bibitem{Berlin:2016woy}
A.~Berlin, \emph{{Neutrino Oscillations as a Probe of Light Scalar Dark
  Matter}}, \href{https://doi.org/10.1103/PhysRevLett.117.231801}{\emph{Phys.
  Rev. Lett.} {\bfseries 117} (2016) 231801}
  [\href{https://arxiv.org/abs/1608.01307}{{\ttfamily 1608.01307}}].

\bibitem{Zhao:2017wmo}
Y.~Zhao, \emph{{Cosmology and time dependent parameters induced by a misaligned
  light scalar}}, \href{https://doi.org/10.1103/PhysRevD.95.115002}{\emph{Phys.
  Rev. D} {\bfseries 95} (2017) 115002}
  [\href{https://arxiv.org/abs/1701.02735}{{\ttfamily 1701.02735}}].

\bibitem{Krnjaic:2017zlz}
G.~Krnjaic, P.A.N.~Machado and L.~Necib, \emph{{Distorted neutrino oscillations
  from time varying cosmic fields}},
  \href{https://doi.org/10.1103/PhysRevD.97.075017}{\emph{Phys. Rev. D}
  {\bfseries 97} (2018) 075017}
  [\href{https://arxiv.org/abs/1705.06740}{{\ttfamily 1705.06740}}].

\bibitem{Brdar:2017kbt}
V.~Brdar, J.~Kopp, J.~Liu, P.~Prass and X.-P.~Wang, \emph{{Fuzzy dark matter
  and nonstandard neutrino interactions}},
  \href{https://doi.org/10.1103/PhysRevD.97.043001}{\emph{Phys. Rev. D}
  {\bfseries 97} (2018) 043001}
  [\href{https://arxiv.org/abs/1705.09455}{{\ttfamily 1705.09455}}].

\bibitem{Davoudiasl:2018hjw}
H.~Davoudiasl, G.~Mohlabeng and M.~Sullivan, \emph{{Galactic Dark Matter
  Population as the Source of Neutrino Masses}},
  \href{https://doi.org/10.1103/PhysRevD.98.021301}{\emph{Phys. Rev. D}
  {\bfseries 98} (2018) 021301}
  [\href{https://arxiv.org/abs/1803.00012}{{\ttfamily 1803.00012}}].

\bibitem{Liao:2018byh}
J.~Liao, D.~Marfatia and K.~Whisnant, \emph{{Light scalar dark matter at
  neutrino oscillation experiments}},
  \href{https://doi.org/10.1007/JHEP04(2018)136}{\emph{JHEP} {\bfseries 04}
  (2018) 136} [\href{https://arxiv.org/abs/1803.01773}{{\ttfamily
  1803.01773}}].

\bibitem{Capozzi:2018bps}
F.~Capozzi, I.M.~Shoemaker and L.~Vecchi, \emph{{Neutrino Oscillations in Dark
  Backgrounds}},
  \href{https://doi.org/10.1088/1475-7516/2018/07/004}{\emph{JCAP} {\bfseries
  07} (2018) 004} [\href{https://arxiv.org/abs/1804.05117}{{\ttfamily
  1804.05117}}].

\bibitem{Huang:2018cwo}
G.-Y.~Huang and N.~Nath, \emph{{Neutrinophilic Axion-Like Dark Matter}},
  \href{https://doi.org/10.1140/epjc/s10052-018-6391-y}{\emph{Eur. Phys. J. C}
  {\bfseries 78} (2018) 922}
  [\href{https://arxiv.org/abs/1809.01111}{{\ttfamily 1809.01111}}].

\bibitem{Farzan:2019yvo}
Y.~Farzan, \emph{{Ultra-light scalar saving the 3 + 1 neutrino scheme from the
  cosmological bounds}},
  \href{https://doi.org/10.1016/j.physletb.2019.134911}{\emph{Phys. Lett. B}
  {\bfseries 797} (2019) 134911}
  [\href{https://arxiv.org/abs/1907.04271}{{\ttfamily 1907.04271}}].

\bibitem{Cline:2019seo}
J.M.~Cline, \emph{{Viable secret neutrino interactions with ultralight dark
  matter}}, \href{https://doi.org/10.1016/j.physletb.2019.135182}{\emph{Phys.
  Lett. B} {\bfseries 802} (2020) 135182}
  [\href{https://arxiv.org/abs/1908.02278}{{\ttfamily 1908.02278}}].

\bibitem{Dev:2020kgz}
A.~Dev, P.A.N.~Machado and P.~Mart\'\i{}nez-Mirav\'e, \emph{{Signatures of
  ultralight dark matter in neutrino oscillation experiments}},
  \href{https://doi.org/10.1007/JHEP01(2021)094}{\emph{JHEP} {\bfseries 01}
  (2021) 094} [\href{https://arxiv.org/abs/2007.03590}{{\ttfamily
  2007.03590}}].

\bibitem{Losada:2021bxx}
M.~Losada, Y.~Nir, G.~Perez and Y.~Shpilman, \emph{{Probing scalar dark matter
  oscillations with neutrino oscillations}},
  \href{https://doi.org/10.1007/JHEP04(2022)030}{\emph{JHEP} {\bfseries 04}
  (2022) 030} [\href{https://arxiv.org/abs/2107.10865}{{\ttfamily
  2107.10865}}].

\bibitem{Huang:2021kam}
G.-y.~Huang and N.~Nath, \emph{{Neutrino meets ultralight dark matter:
  0\ensuremath{\nu}\ensuremath{\beta}\ensuremath{\beta} decay and cosmology}},
  \href{https://doi.org/10.1088/1475-7516/2022/05/034}{\emph{JCAP} {\bfseries
  05} (2022) 034} [\href{https://arxiv.org/abs/2111.08732}{{\ttfamily
  2111.08732}}].

\bibitem{Chun:2021ief}
E.J.~Chun, \emph{{Neutrino Transition in Dark Matter}},
  \href{https://arxiv.org/abs/2112.05057}{{\ttfamily 2112.05057}}.

\bibitem{Dev:2022bae}
A.~Dev, G.~Krnjaic, P.~Machado and H.~Ramani, \emph{{Constraining feeble
  neutrino interactions with ultralight dark matter}},
  \href{https://doi.org/10.1103/PhysRevD.107.035006}{\emph{Phys. Rev. D}
  {\bfseries 107} (2023) 035006}
  [\href{https://arxiv.org/abs/2205.06821}{{\ttfamily 2205.06821}}].

\bibitem{Huang:2022wmz}
G.-y.~Huang, M.~Lindner, P.~Mart\'\i{}nez-Mirav\'e and M.~Sen,
  \emph{{Cosmology-friendly time-varying neutrino masses via the sterile
  neutrino portal}},
  \href{https://doi.org/10.1103/PhysRevD.106.033004}{\emph{Phys. Rev. D}
  {\bfseries 106} (2022) 033004}
  [\href{https://arxiv.org/abs/2205.08431}{{\ttfamily 2205.08431}}].

\bibitem{Losada:2022uvr}
M.~Losada, Y.~Nir, G.~Perez, I.~Savoray and Y.~Shpilman, \emph{{Parametric
  resonance in neutrino oscillations induced by ultra-light dark matter and
  implications for KamLAND and JUNO}},
  \href{https://doi.org/10.1007/JHEP03(2023)032}{\emph{JHEP} {\bfseries 03}
  (2023) 032} [\href{https://arxiv.org/abs/2205.09769}{{\ttfamily
  2205.09769}}].

\bibitem{Guo:2022vxr}
J.~Guo, Y.~He, J.~Liu, X.-P.~Wang and K.-P.~Xie, \emph{{Unveiling time-varying
  signals of ultralight bosonic dark matter at collider and beam dump
  experiments}},
  \href{https://doi.org/10.1038/s42005-023-01350-6}{\emph{Commun. Phys.}
  {\bfseries 6} (2023) 225} [\href{https://arxiv.org/abs/2206.14221}{{\ttfamily
  2206.14221}}].

\bibitem{Brzeminski:2022rkf}
D.~Brzeminski, S.~Das, A.~Hook and C.~Ristow, \emph{{Constraining Vector Dark
  Matter with neutrino experiments}},
  \href{https://doi.org/10.1007/JHEP08(2023)181}{\emph{JHEP} {\bfseries 08}
  (2023) 181} [\href{https://arxiv.org/abs/2212.05073}{{\ttfamily
  2212.05073}}].

\bibitem{Alonso-Alvarez:2023tii}
G.~Alonso-\'Alvarez, K.~Bleau and J.M.~Cline, \emph{{Distortion of neutrino
  oscillations by dark photon dark matter}},
  \href{https://doi.org/10.1103/PhysRevD.107.055045}{\emph{Phys. Rev. D}
  {\bfseries 107} (2023) 055045}
  [\href{https://arxiv.org/abs/2301.04152}{{\ttfamily 2301.04152}}].

\bibitem{Losada:2023zap}
M.~Losada, Y.~Nir, G.~Perez, I.~Savoray and Y.~Shpilman, \emph{{Time dependent
  CP-even and CP-odd signatures of scalar ultralight dark matter in neutrino
  oscillations}},
  \href{https://doi.org/10.1103/PhysRevD.108.055004}{\emph{Phys. Rev. D}
  {\bfseries 108} (2023) 055004}
  [\href{https://arxiv.org/abs/2302.00005}{{\ttfamily 2302.00005}}].

\bibitem{Davoudiasl:2023uiq}
H.~Davoudiasl and P.B.~Denton, \emph{{Sterile neutrino shape shifting caused by
  dark matter}}, \href{https://doi.org/10.1103/PhysRevD.108.035013}{\emph{Phys.
  Rev. D} {\bfseries 108} (2023) 035013}
  [\href{https://arxiv.org/abs/2301.09651}{{\ttfamily 2301.09651}}].

\bibitem{Gherghetta:2023myo}
T.~Gherghetta and A.~Shkerin, \emph{{Probing a local dark matter halo with
  neutrino oscillations}},
  \href{https://doi.org/10.1103/PhysRevD.108.095009}{\emph{Phys. Rev. D}
  {\bfseries 108} (2023) 095009}
  [\href{https://arxiv.org/abs/2305.06441}{{\ttfamily 2305.06441}}].

\bibitem{ChoeJo:2023ffp}
Y.~ChoeJo, Y.~Kim and H.-S.~Lee, \emph{{Dirac-Majorana neutrino type
  oscillation induced by a wave dark matter}},
  \href{https://doi.org/10.1103/PhysRevD.108.095028}{\emph{Phys. Rev. D}
  {\bfseries 108} (2023) 095028}
  [\href{https://arxiv.org/abs/2305.16900}{{\ttfamily 2305.16900}}].

\bibitem{Chen:2023vkq}
Y.~Chen, X.~Xue and V.~Cardoso, \emph{{Black Holes as Neutrino Factories}},
  \href{https://arxiv.org/abs/2308.00741}{{\ttfamily 2308.00741}}.

\bibitem{ChoeJo:2023cnx}
Y.~ChoeJo, K.~Enomoto, Y.~Kim and H.-S.~Lee, \emph{{Second leptogenesis:
  Unraveling the baryon-lepton asymmetry discrepancy}},
  \href{https://doi.org/10.1007/JHEP03(2024)003}{\emph{JHEP} {\bfseries 03}
  (2024) 003} [\href{https://arxiv.org/abs/2311.16672}{{\ttfamily
  2311.16672}}].

\bibitem{Martinez-Mirave:2024dmw}
P.~Mart\'\i{}nez-Mirav\'e, Y.F.~Perez-Gonzalez and M.~Sen, \emph{{Effects of
  neutrino-ultralight dark matter interaction on the cosmic neutrino
  background}}, \href{https://doi.org/10.1103/PhysRevD.110.055005}{\emph{Phys.
  Rev. D} {\bfseries 110} (2024) 055005}
  [\href{https://arxiv.org/abs/2406.01682}{{\ttfamily 2406.01682}}].

\bibitem{ChoeJo:2024wqr}
Y.~ChoeJo, K.~Enomoto, Y.~Kim and H.-S.~Lee, \emph{{Refined approaches in
  second leptogenesis for the baryon-lepton asymmetry discrepancy}},
  \href{https://arxiv.org/abs/2406.19694}{{\ttfamily 2406.19694}}.

\bibitem{Plestid:2024kyy}
R.~Plestid and S.~Tevosyan, \emph{{The cosmology of ultralight scalar dark
  matter coupled to right-handed neutrinos}},
  \href{https://arxiv.org/abs/2409.17396}{{\ttfamily 2409.17396}}.

\bibitem{Lee:2024rdc}
J.-W.~Lee, \emph{{Neutrino mass and ultralight dark matter mass from the Higgs
  mechanism}},  \href{https://arxiv.org/abs/2410.02842}{{\ttfamily
  2410.02842}}.

\bibitem{Lambiase:2025}
G.~Lambiase, T.K.~Poddar and L.~Visinelli, \emph{{Impact of the cosmic neutrino
  background on black hole superradiance}},
  \href{https://arxiv.org/abs/2503.02940}{{\ttfamily 2503.02940}}.

\bibitem{Turner:1983he}
M.S.~Turner, \emph{{Coherent Scalar Field Oscillations in an Expanding
  Universe}}, \href{https://doi.org/10.1103/PhysRevD.28.1243}{\emph{Phys. Rev.
  D} {\bfseries 28} (1983) 1243}.

\bibitem{Das:2020nwc}
S.~Das and E.O.~Nadler, \emph{{Constraints on the epoch of dark matter
  formation from Milky Way satellites}},
  \href{https://doi.org/10.1103/PhysRevD.103.043517}{\emph{Phys. Rev. D}
  {\bfseries 103} (2021) 043517}
  [\href{https://arxiv.org/abs/2010.01137}{{\ttfamily 2010.01137}}].

\bibitem{Coleman:1973jx}
S.R.~Coleman and E.J.~Weinberg, \emph{{Radiative Corrections as the Origin of
  Spontaneous Symmetry Breaking}},
  \href{https://doi.org/10.1103/PhysRevD.7.1888}{\emph{Phys. Rev. D} {\bfseries
  7} (1973) 1888}.

\bibitem{Hall:2009bx}
L.J.~Hall, K.~Jedamzik, J.~March-Russell and S.M.~West, \emph{{Freeze-In
  Production of FIMP Dark Matter}},
  \href{https://doi.org/10.1007/JHEP03(2010)080}{\emph{JHEP} {\bfseries 03}
  (2010) 080} [\href{https://arxiv.org/abs/0911.1120}{{\ttfamily 0911.1120}}].

\bibitem{KamLAND-Zen:2012uen}
{\scshape KamLAND-Zen} collaboration, \emph{{Limits on Majoron-emitting
  double-beta decays of Xe-136 in the KamLAND-Zen experiment}},
  \href{https://doi.org/10.1103/PhysRevC.86.021601}{\emph{Phys. Rev. C}
  {\bfseries 86} (2012) 021601}
  [\href{https://arxiv.org/abs/1205.6372}{{\ttfamily 1205.6372}}].

\bibitem{ParticleDataGroup:2024cfk}
{\scshape Particle Data Group} collaboration, \emph{{Review of particle
  physics}}, \href{https://doi.org/10.1103/PhysRevD.110.030001}{\emph{Phys.
  Rev. D} {\bfseries 110} (2024) 030001}.

\bibitem{Long:2014zva}
A.J.~Long, C.~Lunardini and E.~Sabancilar, \emph{{Detecting non-relativistic
  cosmic neutrinos by capture on tritium: phenomenology and physics
  potential}}, \href{https://doi.org/10.1088/1475-7516/2014/08/038}{\emph{JCAP}
  {\bfseries 08} (2014) 038} [\href{https://arxiv.org/abs/1405.7654}{{\ttfamily
  1405.7654}}].

\bibitem{Betts:2013uya}
S.~Betts et~al., \emph{{Development of a Relic Neutrino Detection Experiment at
  PTOLEMY: Princeton Tritium Observatory for Light, Early-Universe,
  Massive-Neutrino Yield}},  in \emph{{Snowmass 2013}: {Snowmass on the
  Mississippi}}, 7, 2013 [\href{https://arxiv.org/abs/1307.4738}{{\ttfamily
  1307.4738}}].

\bibitem{PTOLEMY:2018jst}
{\scshape PTOLEMY} collaboration, \emph{{PTOLEMY: A Proposal for Thermal Relic
  Detection of Massive Neutrinos and Directional Detection of MeV Dark
  Matter}},  \href{https://arxiv.org/abs/1808.01892}{{\ttfamily 1808.01892}}.

\bibitem{PTOLEMY:2019hkd}
{\scshape PTOLEMY} collaboration, \emph{{Neutrino physics with the PTOLEMY
  project: active neutrino properties and the light sterile case}},
  \href{https://doi.org/10.1088/1475-7516/2019/07/047}{\emph{JCAP} {\bfseries
  07} (2019) 047} [\href{https://arxiv.org/abs/1902.05508}{{\ttfamily
  1902.05508}}].

\bibitem{Perez-Gonzalez:2023llw}
Y.F.~Perez-Gonzalez and M.~Sen, \emph{{From Dirac to Majorana: The cosmic
  neutrino background capture rate in the minimally extended Standard Model}},
  \href{https://doi.org/10.1103/PhysRevD.109.023022}{\emph{Phys. Rev. D}
  {\bfseries 109} (2024) 023022}
  [\href{https://arxiv.org/abs/2308.05147}{{\ttfamily 2308.05147}}].

\bibitem{Kainulainen:1990ds}
K.~Kainulainen, \emph{{Light Singlet Neutrinos and the Primordial
  Nucleosynthesis}},
  \href{https://doi.org/10.1016/0370-2693(90)90054-A}{\emph{Phys. Lett. B}
  {\bfseries 244} (1990) 191}.

\bibitem{Enqvist:1991qj}
K.~Enqvist, K.~Kainulainen and M.J.~Thomson, \emph{{Stringent cosmological
  bounds on inert neutrino mixing}},
  \href{https://doi.org/10.1016/0550-3213(92)90442-E}{\emph{Nucl. Phys. B}
  {\bfseries 373} (1992) 498}.

\bibitem{Hannestad:2012ky}
S.~Hannestad, I.~Tamborra and T.~Tram, \emph{{Thermalisation of light sterile
  neutrinos in the early universe}},
  \href{https://doi.org/10.1088/1475-7516/2012/07/025}{\emph{JCAP} {\bfseries
  07} (2012) 025} [\href{https://arxiv.org/abs/1204.5861}{{\ttfamily
  1204.5861}}].

\bibitem{Vincent:2014rja}
A.C.~Vincent, E.F.~Martinez, P.~Hern\'andez, M.~Lattanzi and O.~Mena,
  \emph{{Revisiting cosmological bounds on sterile neutrinos}},
  \href{https://doi.org/10.1088/1475-7516/2015/04/006}{\emph{JCAP} {\bfseries
  04} (2015) 006} [\href{https://arxiv.org/abs/1408.1956}{{\ttfamily
  1408.1956}}].

\bibitem{Bolton:2019pcu}
P.D.~Bolton, F.F.~Deppisch and P.S.~Bhupal~Dev, \emph{{Neutrinoless double beta
  decay versus other probes of heavy sterile neutrinos}},
  \href{https://doi.org/10.1007/JHEP03(2020)170}{\emph{JHEP} {\bfseries 03}
  (2020) 170} [\href{https://arxiv.org/abs/1912.03058}{{\ttfamily
  1912.03058}}].

\bibitem{Zeldovich:1974uw}
Y.B.~Zeldovich, I.Y.~Kobzarev and L.B.~Okun, \emph{{Cosmological Consequences
  of the Spontaneous Breakdown of Discrete Symmetry}}, {\emph{Zh. Eksp. Teor.
  Fiz.} {\bfseries 67} (1974) 3}.

\bibitem{Sikivie:1982qv}
P.~Sikivie, \emph{{Of Axions, Domain Walls and the Early Universe}},
  \href{https://doi.org/10.1103/PhysRevLett.48.1156}{\emph{Phys. Rev. Lett.}
  {\bfseries 48} (1982) 1156}.

\bibitem{Vilenkin:1984ib}
A.~Vilenkin, \emph{{Cosmic Strings and Domain Walls}},
  \href{https://doi.org/10.1016/0370-1573(85)90033-X}{\emph{Phys. Rept.}
  {\bfseries 121} (1985) 263}.

\bibitem{Lazanu:2015fua}
A.~Lazanu, C.J.A.P.~Martins and E.P.S.~Shellard, \emph{{Contribution of domain
  wall networks to the CMB power spectrum}},
  \href{https://doi.org/10.1016/j.physletb.2015.06.034}{\emph{Phys. Lett. B}
  {\bfseries 747} (2015) 426}
  [\href{https://arxiv.org/abs/1505.03673}{{\ttfamily 1505.03673}}].

\bibitem{Fabris:2000qz}
J.C.~Fabris and S.V.~de~Borba~Goncalves, \emph{{Evolution of perturbations in a
  domain wall cosmology}},
  \href{https://doi.org/10.1590/S0103-97332003000400039}{\emph{Braz. J. Phys.}
  {\bfseries 33} (2003) 834}
  [\href{https://arxiv.org/abs/gr-qc/0010046}{{\ttfamily gr-qc/0010046}}].

\bibitem{Friedland:2002qs}
A.~Friedland, H.~Murayama and M.~Perelstein, \emph{{Domain walls as dark
  energy}}, \href{https://doi.org/10.1103/PhysRevD.67.043519}{\emph{Phys. Rev.
  D} {\bfseries 67} (2003) 043519}
  [\href{https://arxiv.org/abs/astro-ph/0205520}{{\ttfamily
  astro-ph/0205520}}].

\bibitem{Vilenkin:1981zs}
A.~Vilenkin, \emph{{Gravitational Field of Vacuum Domain Walls and Strings}},
  \href{https://doi.org/10.1103/PhysRevD.23.852}{\emph{Phys. Rev. D} {\bfseries
  23} (1981) 852}.

\bibitem{Hiramatsu:2013qaa}
T.~Hiramatsu, M.~Kawasaki and K.~Saikawa, \emph{{On the estimation of
  gravitational wave spectrum from cosmic domain walls}},
  \href{https://doi.org/10.1088/1475-7516/2014/02/031}{\emph{JCAP} {\bfseries
  02} (2014) 031} [\href{https://arxiv.org/abs/1309.5001}{{\ttfamily
  1309.5001}}].

\bibitem{Saikawa:2017hiv}
K.~Saikawa, \emph{{A review of gravitational waves from cosmic domain walls}},
  \href{https://doi.org/10.3390/universe3020040}{\emph{Universe} {\bfseries 3}
  (2017) 40} [\href{https://arxiv.org/abs/1703.02576}{{\ttfamily 1703.02576}}].

\bibitem{Kitajima:2023cek}
N.~Kitajima, J.~Lee, K.~Murai, F.~Takahashi and W.~Yin, \emph{{Gravitational
  waves from domain wall collapse, and application to nanohertz signals with
  QCD-coupled axions}},
  \href{https://doi.org/10.1016/j.physletb.2024.138586}{\emph{Phys. Lett. B}
  {\bfseries 851} (2024) 138586}
  [\href{https://arxiv.org/abs/2306.17146}{{\ttfamily 2306.17146}}].

\bibitem{Lu:2024dzj}
B.-Q.~Lu, \emph{{Scalar-induced gravitational wave from domain wall
  perturbation}},  \href{https://arxiv.org/abs/2412.07677}{{\ttfamily
  2412.07677}}.

\bibitem{Safdi:2014rza}
B.R.~Safdi, M.~Lisanti, J.~Spitz and J.A.~Formaggio, \emph{{Annual Modulation
  of Cosmic Relic Neutrinos}},
  \href{https://doi.org/10.1103/PhysRevD.90.043001}{\emph{Phys. Rev. D}
  {\bfseries 90} (2014) 043001}
  [\href{https://arxiv.org/abs/1404.0680}{{\ttfamily 1404.0680}}].

\end{thebibliography}\endgroup

\end{document}